\documentclass[twocolumn,aps,prl,superscriptaddress,showpacs,amsmath,amssymb,floats]{revtex4}

\usepackage{amsmath}
\usepackage{amssymb}
\usepackage{graphicx}
\usepackage{CJK}
\usepackage{keyval}
\usepackage{dcolumn}
\usepackage{bm}
\usepackage{color}
\usepackage{txfonts}
\usepackage[]{sidecap}

\begin{document}

\begin{CJK*}{UTF8}{gbsn}

\title{Temperature-induced band selective localization and coherent-incoherent crossover in single-layer FeSe/Nb:BaTiO$_3$/KTaO$_3$}

\author{Y. J. Pu}\author{Z. C. Huang}\author{H. C. Xu}\author{D. F. Xu}\author{Q. Song}\author{C. H. P. Wen}

\affiliation{State Key Laboratory of Surface Physics, Department of Physics, and Laboratory of Advanced Materials, Fudan University, Shanghai 200433, China}
\affiliation{Collaborative Innovation Center of Advanced Microstructures, Nanjing 210093, China}
\author{R. Peng}\email{pengrui@fudan.edu.cn}\author{D. L. Feng}\email{dlfeng@fudan.edu.cn}
\affiliation{State Key Laboratory of Surface Physics, Department of Physics, and Laboratory of Advanced Materials, Fudan University, Shanghai 200433, China}
\affiliation{Collaborative Innovation Center of Advanced Microstructures, Nanjing 210093, China}
\date{\today}
\begin{abstract}
Iron chalcogenide superconductors are multi-band systems with strong electron correlations. Here we use angle-resolved photoemission spectroscopy to study band dependent correlation effects in single-layer FeSe/Nb:BaTiO$_3$/KTaO$_3$, a new iron chalcogenide superconductor with non-degenerate electron pockets and interface-enhanced superconductivity. The non-degeneracy of the electron bands helps to resolve the temperature dependent evolution of different bands. With increasing temperature, the single layer FeSe undergoes a band-selective localization, in which the coherent spectral weight of one electron band is completely depleted while that of the other one remains finite. In addition, the spectral weight of the incoherent background is enhanced with increasing temperature, indicating a coherent-incoherent crossover. Signatures of polaronic behavior are observed, suggesting electron-boson interactions. These phenomena help to construct a more complete picture of electron correlations in the FeSe family.
\end{abstract}


\maketitle

\end{CJK*}
\section{I.  Introduction}

Understanding the electron correlations is a central issue in studying unconventional superconductivity. Iron pnictide superconductors show moderate electron correlations with a metallic parental phase, and thus weak coupling theories suggest that the half filled Mott-Hubbard model used in the cuprates may not be a good starting point for studying their superconductivity \cite{Hirschfeld}. On the other hand, there is increasing evidence suggesting that iron chalcogenide superconductors are rather strongly correlated, which has attracted intense research interests\cite{YZP,FeSe_strongCorrelation,FeTeSe_strongCorrelation,FeSe_bulk,LZK_FeTeSe,wenchen,KFeSeTe,FeSe_mott,KFeSe_mott,(LiFe)OHFeSe,Phase separation,ZKLiu0,ZKLiu}.
Especially, an orbital selective Mott phase has been predicted to emerge due to the cooperative effect of Hund's rule coupling and the on-site Coulomb repulsion \cite{OSMT1,OSMT2,SQM1,SQM2,1_concept_OSMT}. To get a unified understanding on the correlation effects in Fe-based superconductors, it is crucial to experimentally explore the orbital selective Mott phase and to resolve its characteristics in the electronic structure.

Angle resolved photoemission spectroscopy (ARPES) can directly probe the behavior of orbital selective Mott transition by resolving the correlation behavior of different bands and orbitals. Pioneering ARPES studies by M. Yi $et~al.$ have shown signatures of temperature induced Mott crossovers for the d$_{xy}$ orbital between 100 K and 200 K in $A$$_x$Fe$_{2-y}$Se$_2$($A$=K, Rb), FeTe$_{0.56}$Se$_{0.44}$, and single-layer FeSe/SrTiO$_3$ \cite{ZX1,ZX2}. Nevertheless, in these materials, the d$_{xy}$ bands are rather weak or nearly degenerate with other bands near the Fermi energy (E$_F$), which complicates the analysis on the spectral weight evolution of different orbitals. The previous analysis was based on the temperature dependent evolution in the energy distribution curves (EDCs), where the spectral weight of d$_{xy}$ bands diminishes at high temperatures, which cannot be explained by the thermal broadening effect \cite{ZX1,ZX2}.
However, anomalous temperature dependent evolution beyond thermal broadening is common in the EDCs of strongly correlated systems \cite{Kim},
which would interfere the temperature dependence of an individual band.
Therefore, exploring orbital selective Mott physics requires ARPES studies on more iron chalcogenide materials, especially those with well separated d$_{xy}$ bands.

Recently, single layer FeSe films have been successfully grown on BaTiO$_3$, which can show two types of domains based on the growth conditions, including rotated and unrotated domains named as FeSe$^{BR}$ phase and FeSe$^{BU}$ phase, respectively \cite{BTO}. Among these two domains, the unrotated domain FeSe$^{BU}$ shows a pairing temperature of 75~K, the highest among all Fe-based superconductors based on in-situ ARPES study \cite{BTO}. Moreover, the two electron bands around the Brillouin zone corner in FeSe$^{BU}$ are well separated in momentum space, and thus the temperature dependence of the spectral weight of these two bands can be well resolved by momentum distribution curves (MDCs) instead of EDCs, which would avoid the complexities from the broadening of EDCs induced by many-body interactions. Therefore, it serves as a new and ideal material to explore the possible orbital selective Mott physics in iron chalcogenides.

In this paper, we have performed systematic ARPES studies on the temperature dependent electronic structure of FeSe$^{BU}$. With increasing temperature, significant broadening has been observed in energy, which is beyond the thermal broadening. Nevertheless, the temperature dependent studies on the MDCs resolve that the coherent spectral weight of one electron band is depleted at 200~K, while that of the other electron band remains finite, giving direct evidence of a band selective localization. Moreover, the incoherent spectral weight at $E_F$ significantly increases on increasing temperature, suggesting a coherent-incoherent crossover. These results provide explicit evidence on strong electron correlations and orbital selective Mott physics in iron chalcogenides.

\section{II. EXPERIMENTAL DETAILS}
\begin{figure}[tb]
\includegraphics[width=8.6cm]{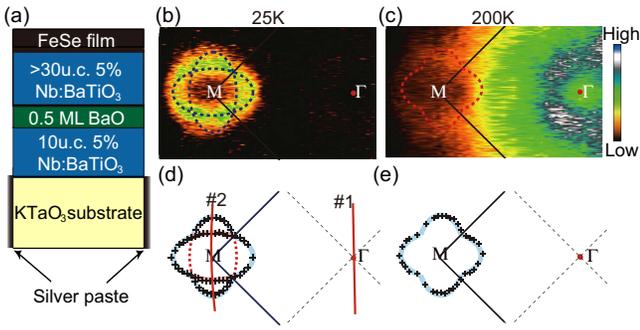}
\caption{(color online).
(a) The schematic cross-section of the FeSe$^{BU}$ film.
(b-c) Photoemission intensity maps at $E_F$ of FeSe$^{BU}$ measured at 25~K and 200~K, respectively.
(d) The local maxima in the MDCs at $E_F$ (cross marks in black) and the illustration of the Fermi surfaces by tracking the Fermi crossings at $E_F$ at 25~K (a quatrefoil shaped outer Fermi surface indicated by the light blue solid line and a square shaped inner Fermi surface indicated by the red solid and dashed lines). Although the parts indicated by the red dashed line are ambiguous due to matrix element effects, we can obtain it based on the symmetry. (e)
The local maxima in the MDCs at $E_F$ (cross marks in black) and the illustration of the Fermi surfaces by tracking the Fermi crossings at $E_F$ at 200~K (Fermi surface indicated by the light blue solid line)}
\label{map}
\end{figure}

The single layer FeSe films with pure FeSe$^{BU}$ phase were grown following our previous report \cite{BTO}, with the schematic cross-section shown in Fig.~\ref{map}(a).
Before growth, KTaO$_3$(001) substrates were etched with buffered NH$_4$-HF solution and annealed in 2~bar oxygen to obtain atomically flat surfaces. To avoid the photoemission charging effect of insulating KTaO$_3$, silver paste was attached on the substrate edge. The Nb:BaTiO$_3$ thin films with 5\%~Nb doping were grown in atomic layer-by-layer mode with ozone-assisted molecular beam epitaxy \cite{BTO}. In order to get pure FeSe$^{BU}$, half a layer of BaO was inserted during the growth of the Nb:BaTiO$_3$ thin films \cite{BTO}.
The grown Nb:BaTiO$_3$/KTaO$_3$ heterostructures were transferred under ultra-high vacuum to another MBE chamber where single layer FeSe thin films were grown \cite{BTO,tan}. After growth, the samples were transferred under ultrahigh vacuum for in-situ ARPES measurements. ARPES data were taken under ultra-high vacuum of 1.5$\times$ 10$^{-11}$mbar, with a SPECS UVLS discharge lamp (21.2eV He-I$\alpha$ photons) and a Scienta R4000 electron analyzer. The overall energy resolution is 8 meV and the angular resolution is $0.3^{\circ}$.

\begin{figure}[tb]
\includegraphics[width=8.6cm]{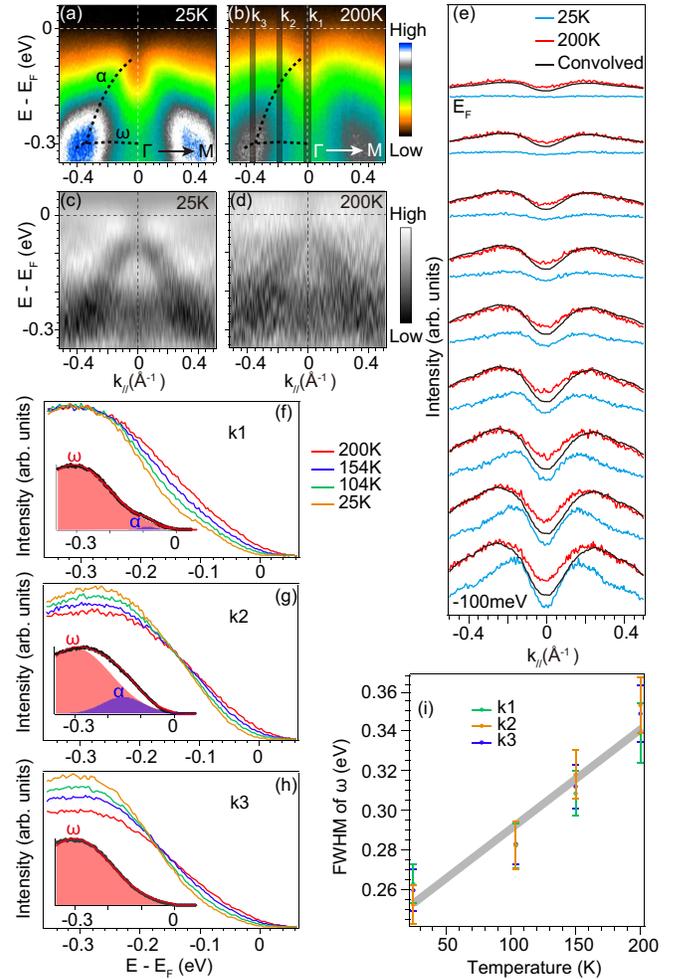}
\caption{(color online).
(a-b) Photoemission intensity around $\Gamma$ [cut \#1 in Fig.~\ref{map}(d)] at 25~K and 200~K, respectively. The dashed lines illustrate the dispersion of the $\alpha$ and $\omega$ bands. 
(c-d) The second derivative with respect to energy of the data in (a) and (b).
(e) The MDCs of spectra at 200~K (red curves), 25~K (blue curves), and the 25~K spectra convolved by a Gaussian function with the full width at half maximum (FWHM) of 210~meV (black curves).
(f-h) The temperature dependent evolution of the EDCs integrated near momenta $k_1$ (f), $k_2$ (g), and $k_3$ (h) as indicated in panel (b), respectively. The insets show the fitting result of the 25~K EDCs by Gaussian peaks corresponding to the $\alpha$ and $\omega$ bands.
(i) The temperature dependence of the FWHM of the $\omega$ band at momenta $k_1$, $k_2$, and $k_3$.
}
\label{Gamma}
\end{figure}

\begin{figure*}
\includegraphics[width=18cm]{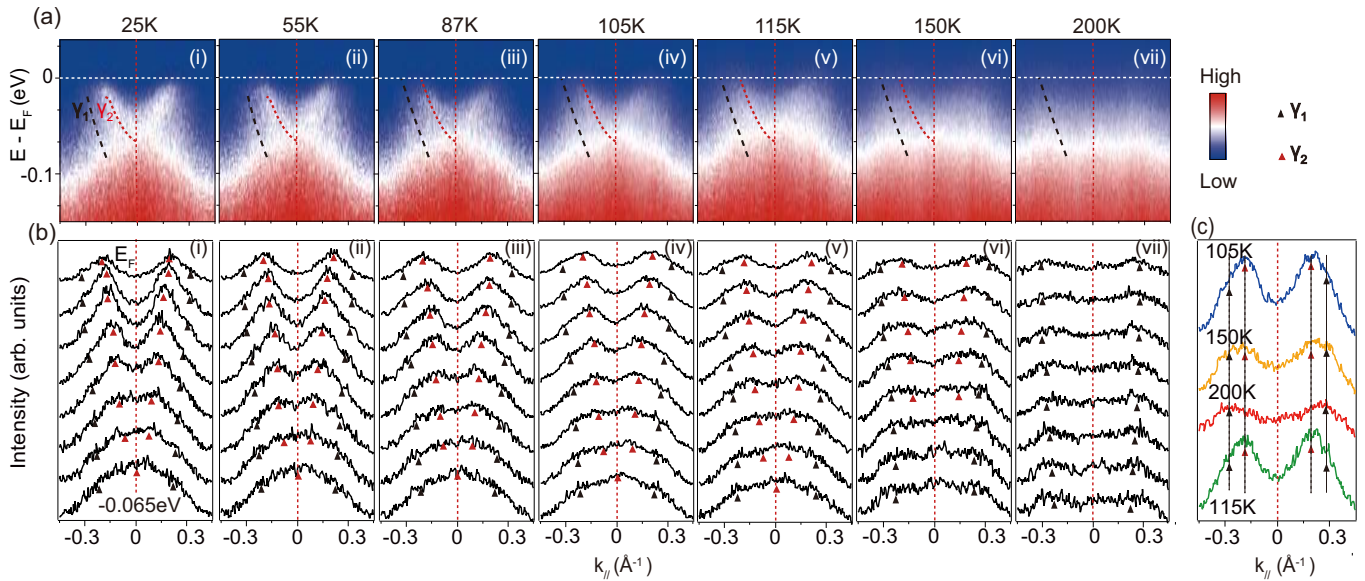}
\caption{(color online)
(a) Photoemission spectra around M [cut \#2 in Fig.~\ref{map}(d)] as a function of temperature.
(b) Corresponding MDCs between [E$_F$$-$65meV, E$_F$] of the photoemission spectra in panels (a). The band dispersion is tracked by local maxima in the MDCs and sketched by triangle marks.
(c) Temperature dependence of the MDCs at E$_F$ from 105~K to 200~K, and a cycle back to 115~K.
}
\label{M}
\end{figure*}

\begin{figure}[tb]
\includegraphics[width=8.6cm]{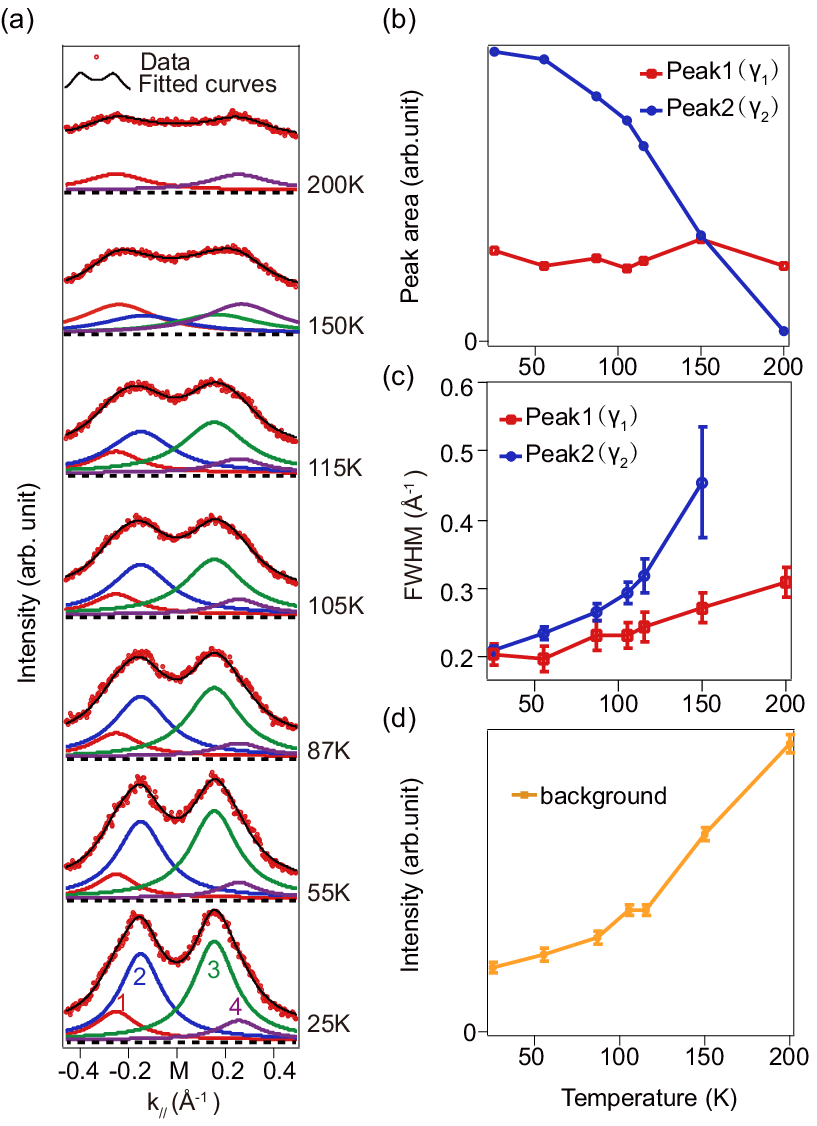}
\caption{(color online)
(a) Temperature dependence of the MDCs integrated over [E$_F$$-$30 meV, E$_F$$-$15 meV] around M, and the corresponding fittings by Lorentzian peaks and constant backgrounds. Peak1 and peak4 (peak2 and peak3) correspond to the peak positions of the $\gamma_1$ ($\gamma_2$) band. The MDCs and fitted curves are vertically offset for clarity, and the horizontal dashed lines indicate the zero intensity for the MDCs at each temperature. (b) The evolution of the peak areas of peak1 and peak2 in panel a, indicating the spectral weight for different bands. The peak areas are obtained by integrating the intensity of the Lorentzian peaks over its momentum section.
(c) FWHM of peak1 and peak2 as a function of temperature.
(d) Intensity of the constant background as a function of temperature.
}
\label{DOS}
\end{figure}

\section{III. RESULTS}

Based on the photoemission intensity maps at $E_F$, the Fermi surfaces of FeSe$^{BU}$ at 25~K consist of two elliptical pockets around M and there is no spectral weight at E$_F$ around $\Gamma$ [Fig.~\ref{map}(b)], which is consistent with our previous report \cite{BTO}. The two elliptical pockets form a quatrefoil shaped outer Fermi surface and a square shaped inner Fermi surface around M [Fig.~\ref{map}(d)]. Although the parts parallel to the direction of the cut indicated by the red dashed line are ambiguous due to matrix element effects [Fig.~\ref{map}(d)], we can obtain it based on the symmetry. Intriguingly, the Fermi surface map looks significantly different at 200~K.
The spectral weight at $E_F$ is enhanced over the entire Brillouin zone [Fig.~\ref{map}(c)].
Although the spectral weight at $E_F$ is momentum dependent and forms an annular shape around $\Gamma$ [Fig.~\ref{map}(c)], it is very broad and incoherent without forming a sharp Fermi surface, which will be further discussed in Fig.~\ref{Gamma}. The Fermi pockets around M are barely visible due to the overwhelming contribution of the incoherent spectral weight [Fig.~\ref{map}(c)]. Nevertheless, by tracking the local maxima of MDCs, one could observe a quatrefoil shaped Fermi surface around M [Fig.~\ref{map}(e)], which corresponds to the outer part of the Fermi surface at 25~K [Fig.~\ref{map}(d)]. The square shaped, inner part of the Fermi surface around M is not visible at 200~K [Fig.~\ref{map}(e)]. The dramatic evolution of the Fermi surfaces indicates possible band structure reconstruction with increasing temperature.

To study this, we first inspect the temperature dependence of spectra around $\Gamma$. At 25~K, there is no spectral weight at $E_F$ around $\Gamma$, and there are a parabolic band $\alpha$ and a flat band $\omega$ below E$_F$ [Fig.~\ref{Gamma}(a)], consistent with the previous report \cite{BTO}.
Intriguingly, at 200~K, dispersive features emerge at E$_F$, as shown by the MDCs in Fig.~\ref{Gamma}(e), which form the annular shaped spectral weight maxima around $\Gamma$ [Figs.~\ref{map}(c)].
However, the second derivatives of the spectra at 200~K and 25~K with respect to energy show similar band dispersions [Figs.~\ref{Gamma}(c) and \ref{Gamma}(d)], indicating no reconstruction of the band structure around $\Gamma$ as a function of temperature.

At 200~K, all the features appear broadened compared with those at 25~K [Fig.~\ref{Gamma}(b)]. The EDCs at various momenta can be well fitted by Gaussian peaks representing the bands $\alpha$ and $\omega$ [insets of Figs.~\ref{Gamma}(f)-\ref{Gamma}(h)]. When the temperature increases from 25~K to 200~K, the FWHM of the $\omega$ band increases from 260~meV to 340~meV for all the momenta [Fig.~\ref{Gamma}(i)], suggesting a momentum independent broadening of $\sqrt{340^2-260^2}$~meV=219~meV in energy. As shown in Fig.~\ref{Gamma}(i), the broadening occurs over an extended temperature range following an almost constant rate of around 0.5$\pm$0.08~meV/K.
Remarkably, if we manually broaden the low temperature spectrum by convolving a Gaussian function with FWHM=210~meV in energy, and then multiply it by the Fermi-Dirac function of 200~K, the resulting spectrum can well reproduce the spectrum of 200~K as shown by the MDCs near E$_F$ in Fig.~\ref{Gamma}(e). Thus the emergent spectral weight at $E_F$ around $\Gamma$ at high temperature basically comes from the broadening of bands below $E_F$.
The energy broadening of $\sim$210~meV from 25~K to 200~K is significantly larger than the thermal broadening, which is $4k_B{\Delta}T$=60~meV. Therefore, this temperature dependent broadening is a non-trivial effect with an energy scale far exceeding that of the thermal broadening.

At 25~K, the photoemission spectra around M show two electron-like bands $\gamma_1$ and $\gamma_2$ crossing E$_F$ [Fig.~\ref{M}(a)(i)], which form the outer and inner Fermi surface sections around M, respectively [Figs.~\ref{map}(b),(d)].
With increasing temperature, although the band dispersions do not notably change, the $\gamma_2$ band gradually weakens in intensity, and disappears entirely by 200~K, while the $\gamma_1$ band remains [Figs.~\ref{M}(a),(b)]. The loss of spectral weight of the $\gamma$$_2$ band occurs gradually over a wide range of temperature, indicating a temperature induced crossover rather than a sharp transition.
The spectral weight of the $\gamma$$_2$ band is recovered when the temperature is cycled back [Figs.~\ref{M}(c)], demonstrating that its temperature dependent evolution is intrinsic.

To quantify the temperature dependent evolution near M, we fit the MDCs by Lorentzian peaks [Fig.~\ref{DOS}(a)]. As shown in Fig.~\ref{DOS}(b), the spectral weight of the $\gamma$$_2$ band drops much faster than that of the $\gamma$$_1$ band. The spectral weight is totally depleted at 200~K for the $\gamma$$_2$ band, while finite intensity remains for the the $\gamma$$_1$ band, indicating band selective localization. As the temperature rises from 25~K to 200~K, the FWHM of the $\gamma$$_2$ band in momentum significantly increases, while that of the $\gamma$$_1$ band only increases slightly [Fig.~\ref{DOS}(c)]. Meanwhile, the intensity of the momentum independent background is significantly enhanced [Fig.~\ref{DOS}(d)], which may be partly contributed by the spectral weight of the $\gamma$$_2$ band losing coherence. These quantitative results not only provide direct evidence on the band selective localization, but also clearly resolve the electronic evolution during the localization process.

\section{IV.  DISCUSSION AND CONCLUSION }

In the temperature dependent ARPES spectra of FeSe$^{BU}$, the EDCs are significantly broadened with increasing temperature.
Their Gaussian lineshape, broadening over an wide temperature range, is similar to the character of Franck-Condon broadening as observed in polaronic systems \cite{75,87}. Moreover, the EDCs broaden at a rate of $~$0.5$\pm$0.08~meV/K, which is the same order of magnitude as the 0.3~meV/K reported in the iridate Sr$_3$Ir$_2$O$_7$, or 1.0$\pm$0.3~meV/K for the lower Hubbard band, 0.6$\pm$0.4~meV/K for the O~2$p$$_\pi$, and 0.7$\pm$0.2~meV/K for the Ca~3$p$ core level in the cuprate Ca$_2$CuO$_2$Cl$_2$ \cite{75,87}. 
These signatures of polaronic behavior may indicate strong interactions between electrons and phonons or the spin fluctuation in FeSe$^{BU}$. 
It has been suggested that electon-phonon and magnetic interactions may enhance the superconducting pairing and lead to the high temperature superconductivity in single layer FeSe on oxide interface \cite{LeeJJ,BTO}.
Nevertheless, in FeSe$^{BU}$, the non-trivial broadening is observed in the nearly flat band $\omega$ and the renormalized band $\alpha$, which is intriguingly different from the reported polaronic systems where the broad incoherent feature follows bare band dispersion without renormalization \cite{75,87}. 
A comprehensive explanation of the broadening in energy with increasing temperature in FeSe$^{BU}$ and its relation with polaronic physics call for further investigation.

Band selective localization is directly observed in FeSe$^{BU}$, where the $\gamma_2$ band vanishes at 200~K in FeSe$^{BU}$, while the $\gamma_1$ band remains metallic. Since $\gamma_2$ is shallower than $\gamma_1$ with a larger band mass, the $\gamma_2$ band should be comprised of d$_{xy}$ orbital based on the previous ARPES studies on other heavily electron doped iron chalcogenides \cite{ZX1,ZX2}. In this case, the observed band selective behavior with increasing temperature is consistent with the orbital selective Mott crossover observed in A$_x$Fe$_{2-y}$Se$_2$ (A=K, Rb) \cite{ZX1,OS1,OS2,OS3}, single-layer FeSe/SrTiO$_3$ and FeTe$_{0.56}$Se$_{0.44}$ \cite{ZX2}. Our observation extends the presence of orbital selective Mott crossover to a new iron chalcogenide material. Moreover, band/orbital dependent correlation behavior is observed in the momentum-resolved perspective for the first time, which excludes the complexity introduced by broadening in energy, thus giving direct evidence of orbital selective Mott physics.

Instead of band narrowing, the temperature induced localization of the $\gamma_2$ band is manifested as the fading out of the coherent spectral weight, the increment of the FWHM in momentum, and the enhancement of the incoherent background. These characteristics are consistent with the coherent-incoherent crossover in the theories of the correlated metals near an orbital-selective Mott transition \cite{Hardy,ZPYinPRB2012}. With Hund's coupling induced strong correlations, it is predicted that strong incoherence happens at high temperature with some electronic bands fading out in ARPES spectra, while these bands become coherent and appear at low temperatures \cite{Hardy,ZPYinPRB2012}, which is explicitly reflected in our data.

To summarize, we have resolved the temperature dependent correlation effects in FeSe$^{BU}$. With increasing temperature, significant broadening is observed in energy. Moreover, FeSe$^{BU}$ undergoes a band-selective localization, and the non-degeneracy of the electron bands allows us to quantitatively resolve the electronic evolution of different bands. The localization process is observed as a coherent-incoherent crossover, providing a direct experimental basis and quantitative details for understanding orbital selective Mott physics. These results help to unveil the strong electron correlations in the FeSe system and also shed light on other multi-orbital correlated materials.

\section{ACKNOWLEDGEMENTS}
We gratefully acknowledge Prof. George Sawatzky, Prof. Dunghai Lee and Dr Darren Peets for helpful discussions. This work is supported in part by the National Science Foundation of China and the National Basic Research Program of China (973 Program) under the grant No. 2012CB921402, and Science and Technology Commission of Shanghai Municipality under the grant No. 15ZR1402900.

\end{document}